%% file: QAP.tex
\documentclass{article}

\usepackage[round,comma]{natbib}
\usepackage{fullpage}

\usepackage[utf8]{inputenc} 
\usepackage[T1]{fontenc}    
\usepackage{hyperref}       
\usepackage{url}            
\usepackage{booktabs}       
\usepackage{amsfonts}       
\usepackage{nicefrac}       
\usepackage{microtype}      
\usepackage[colorinlistoftodos, textwidth=26mm, shadow,color=blue!30!white]{todonotes}
\usepackage{wrapfig}

\usepackage{subcaption}

\usepackage{algorithm,algorithmic}
\usepackage{verbatim}

\usepackage{amsmath,amsthm,amssymb,amsfonts}

\input{Commands}

\allowdisplaybreaks

\newcommand{\DIV}{\nabla\cdot}
\newcommand{\curl}{\nabla\times}

\title{Fast Mixing Random Walks and Regularity of Incompressible Vector Fields}

\author{Yasin Abbasi-Yadkori}

\begin{document}

\maketitle

\begin{abstract}
We show sufficient conditions under which the \textsc{BallWalk} algorithm mixes fast in a bounded connected subset of $\Real^n$. In particular, we show fast mixing if the space is the transformation of a convex space under a smooth incompressible flow. 
Construction of such smooth flows is in turn reduced to the study of the regularity of the solution of the Dirichlet problem for Laplace's equation. 
\end{abstract}

\input{Introduction}

\input{laplace}













\bibliography{biblio}


\end{document}

%% file: Commands.tex
\usepackage{pdfsync}
\usepackage{graphicx} 

\usepackage{algorithm}
\usepackage{algorithmic}

\usepackage{amssymb}

\usepackage{amsmath} 

\usepackage{color}
\usepackage{graphicx}
\usepackage{epstopdf}
\usepackage{algorithm,algorithmic}
\usepackage{verbatim}

\usepackage{nicefrac}

\newif\ifcomm
\commtrue

\newif\iflong
\longtrue

\newtheorem{thm}{Theorem}

\newtheorem{lem}[thm]{Lemma}

\newtheorem{defn}{Definition}

\newcounter{assumption}
\renewcommand{\theassumption}{A\arabic{assumption}}

\newcommand{\vol}{\textsc{vol}}

\newcommand{\norm}[1]{\left\Vert#1\right\Vert}

\newcommand{\abs}[1]{\left\vert#1\right\vert}

\newcommand{\trace}{\mathop{\rm trace}}

\newcommand{\Real}{\mathbb R}                        

\newcommand{\ra}{\rightarrow}

\newcommand{\argmin}{\mathop{\rm argmin}}

\newcommand{\beq}{\begin{equation}}
\newcommand{\eeq}{\end{equation}}
\newcommand{\beqa}{\begin{eqnarray}}
\newcommand{\eeqa}{\end{eqnarray}}
\newcommand{\beqan}{\begin{eqnarray*}}
\newcommand{\eeqan}{\end{eqnarray*}}
\newcommand{\ben}{\begin{eqnarray*}}
\newcommand{\een}{\end{eqnarray*}}
\newcommand{\bea}{\begin{align*}}
\newcommand{\eea}{\end{align*}}

\ifcomm
   \newcommand\comm[1]{\textcolor{blue}{ #1}}
   \newcommand{\mtodo}[2]{\todo{{\bf #1}: #2}} 
   \def\here#1{{\bf $\langle\langle$#1$\rangle\rangle$}}

\else
   \newcommand\comm[1]{}
   \newcommand{\mtodo}[2]{}
   \def\here#1{}
\fi

\renewcommand{\phi}{\varphi}



%% file: Introduction.tex

\section{Introduction}

A number of random walk methods are known to have fast mixing rates in convex spaces. We are interested in sampling from a bounded connected space $\Omega'\subset\Real^n$ that might be non-convex. Such sampling methods can be useful in solving general optimization and planning problems. Recently, \citet{Abbasi-Bartlett-Gabillon-Malek} analyzed the \textsc{Hit-n-Run} algorithm under the assumptions that (i) a biLipschitz measure-preserving mapping between a convex space $\Omega$ and the target space $\Omega'$ exists, and (ii) $\Omega'$ has low curvature. In this paper, we construct such mappings. Further, we show that existence of such mappings is sufficient to have fast mixing for another popular random walk known as the \textsc{BallWalk}. Thus the curvature condition is not needed to analyze the \textsc{BallWalk} algorithm.


A popular approach to analyze mixing rates is by showing lower bounds for {\em conductance}, which is usually obtained by establishing an isoperimetric inequality. As an example of an isoperimetric inequality, \citet{Dyer-Frieze-1991} show that for any partition $(\Omega_1,\Omega_2,\Omega_3)$ of a convex unit volume $\Omega$, 
\beq
\label{eq:iso-ineq}
\vol(\Omega_3) \ge \frac{2 d(\Omega_1,\Omega_2)}{D_{\Omega}} \min(\vol(\Omega_1), \vol(\Omega_2)) \;.
\eeq
Here $\vol$ denotes the $n$-dimensional volume,\footnote{We will use $\vol_{n-1}$ to denote the  $(n-1)$-dimensional volume.} $D_{\Omega}$ denotes the diameter of $\Omega$ obtained by $D_{\Omega} = \max_{x,y\in \Omega} \abs{x-y}$ where $\abs{x-y}$ is the Euclidean distance between $x,y\in\Real^n$, and $d(\Omega_1,\Omega_2)=\min_{x\in \Omega_1, y\in \Omega_2} \abs{x-y}$. The only isoperimetric inequality for a non-convex space is shown by \citet{Chandrasekaran-Dadush-Vempala-2010} that obtains an inequality for star-shaped bodies.\footnote{We say $\Omega'$ is star-shaped if the kernel of $\Omega'$, define by $N_S = \{x\in \Omega' : \forall y\in \Omega'\,\, [x,y]\subset \Omega' \}$, is nonempty.} 
In this paper, we show that an isoperimetic inequality holds for any non-convex space, as long as there exists a smooth measure-preserving mapping from a convex space to that non-convex space. We also show a construction for such mappings by constructing appropriate smooth incompressible flows. Given such mapping, we show a polynomial mixing rate for \textsc{BallWalk}. 

The \textsc{BallWalk} algorithm is a simple random walk procedure and is defined as follows. Let $B(x,r)$ denote the $n$-dimensional Euclidean ball of radius $r$ centered around $x$. At time $t$, we pick a point uniformly at random from $B(x_t,r)$. Let this point be $y_t$. If the new point is outside $\Omega'$, the move is rejected and $x_{t+1}=x_t$. Otherwise, $x_{t+1} = y_t$. \citet{Kannan-Lovasz-Simonovits-1997} show a polynomial mixing rate for the \textsc{BallWalk} algorithm when $\Omega'$ is convex. 
The next theorem shows a more general result under an embedding assumption. Before giving more details, let us define some notation. Let $\Omega$ be a convex space with boundary $\partial \Omega$, and let $\Omega'$ be a bounded connected subset of $\Real^n$. Assume $\Omega'$ is the image of $\Omega$ under a Lipschitz measure-preserving mapping $g:\Real^n \ra \Real^n$: 
\[
\exists L_{\Omega'}>0, \forall x,y\in \Omega,\quad \abs{g(x)- g(y)} \le L_{\Omega'} \abs{x-y}, \quad \det(D_x g) = 1 \;.
\]
Here $D_x g$ is an $n\times n$ matrix whose $(i,j)$-th element is $\partial g_i / \partial x_j$, also known as the Jacobian matrix evaluated at point $x$. Mapping $g$ is called measure-preserving if $\det(D_x g)=1$ for all $x\in\Omega$. Let $R_{\Omega'}$ be an upper bound on the isoperimetric ratio of $\Omega'$, i.e. $\vol_{n-1}(\partial \Omega)/\vol(\Omega) \le R_{\Omega'}$. 
\begin{thm}
\label{thm:sampling:main}
Consider the \textsc{BallWalk} algorithm. Let $\sigma_0$ be the distribution of the initial point, $\sigma_t$ be the distribution after $t$ steps of the random walk, and $\sigma$ be the uniform distribution on $\Omega'$. Suppose there exist $M>0$ such that for any $A\subset \Omega'$, $\sigma_0(A)\le M \sigma(A)$. For any $\epsilon>0$, after $O((1/\epsilon^2)\log (1/\epsilon))$ steps of the \textsc{BallWalk} with radius $r=O(\epsilon)$, we have $d_{tv}(\sigma_t, \sigma) \le \epsilon$. Here, $d_{tv}$ denotes the total variation distance, and big-O notation hides constants and polynomial terms in $D_\Omega, L_{\Omega'}, R_{\Omega'}, M, n$. 
\end{thm}
Here we show a proof sketch for Theorem~\ref{thm:sampling:main}. First we show that an isoperimetric inequality can be obtained from the embedding assumption. We will discuss existence of such embeddings in the next section. Let $(\Omega_1,\Omega_2,\Omega_3)$ be a partition of $\Omega$ and $(\Omega_1',\Omega_2',\Omega_3')$ be the corresponding partition of $\Omega'$ under mapping $g$. 
Let $(x_1,x_2) = \argmin_{y_1\in \Omega_1, y_2\in \Omega_2} d(y_1,y_2)$. By the embedding assumption,  
\[
d(\Omega_1,\Omega_2) = d(x_1,x_2) \ge \frac{1}{L_{\Omega'}}d(g(x_1), g(x_2)) \ge \frac{1}{L_{\Omega'}} d(\Omega_1', \Omega_2') \;.
\]
Thus,
\begin{align*}
\vol(\Omega_3') = \vol(\Omega_3) &\ge  \frac{1}{4 D_\Omega} d(\Omega_1,\Omega_2) \min\{\vol(\Omega_1), \vol(\Omega_2) \} &\dots \text{ By \eqref{eq:iso-ineq},} \\
&\ge \frac{1}{4 D_\Omega L_{\Omega'}} d(\Omega_1', \Omega_2') \min\{ \vol(\Omega_1'), \vol(\Omega_2')\} &\dots\text{ By embedding assumption.}
\end{align*}
It can be shown that if an isoperimetric inequality holds for $\Omega'$, the Markov process induced by the \textsc{BallWalk} algorithm has large conductance. This step of the proof is omitted and is based on standard techniques from the literature on random walk analysis~\citep{Vempala-2005, Chandrasekaran-Dadush-Vempala-2010}. Next we show the relationship between conductance and mixing rates. Let $P_x(A)$ be the probability of being in set $A\subset \Omega'$ after one step of the process that starts from $x$. The ergodic flow of Markov process is defined by $\Phi(A) = \int_A P_x(\Omega'\setminus A)dx$. Define the $s$-conductance (for $0\le s \le 1$) of the Markov process by
\[
\Phi_s = \inf_{s < \sigma(A) \le 1/2} \frac{\Phi(A)}{\min\{\vol(A), \vol(\Omega'\setminus A)\}} \;.
\]
The following standard result relates the $s$-conductance to the mixing rate.
\begin{lem}[Corollary~1.5 of \citet{Lovasz-Simonovits-1993}]
\label{lem:convergence}
Let $0<s\le 1/2$ and $H_s = \sup_{\sigma(A) \le s} \abs{\sigma_0(A)-\sigma(A)}$. Then for every measurable $A\subset \Real^n$ and every $t\ge 0$,
\[
\abs{\sigma_t(A)-\sigma(A)} \le H_s + \frac{H_s}{s} \left( 1 - \frac{\Phi_s^2}{2} \right)^t \;.
\]
\end{lem}
The lemma shows that the mixing time of the Markov process is directly related to its $s$-conductance. The rest of the paper is devoted to construction of Lipschitz measure-preserving mappings. First we define some notation. 


\subsection{Notation}

We use $\nabla f$, $\nabla^2 f$, $\DIV v$, to denote the gradient of function $f$, Laplacian of $f$, and divergence of vector field $v$, respectively. For integer $k\ge 0$ and $0<\alpha\le 1$, we use $C^{k,\alpha}(K,K')$ to denote the H\"{o}older space, i.e. the space containing mappings from $K$ to $K'$ that have continuous derivatives up to order $k$ and such that the $k$th partial derivatives are H\"{o}lder continuous with exponent $\alpha$. Further, we use $\norm{.}_{k,\alpha}$ to denote the $C^{k,\alpha}$ norm. For integer $k\ge 0$ and $p\ge 1$, we use $W^{k,p}(K,K')$ to denote the Sobolev space.

%% file: laplace.tex

\section{Lipschitz Measure-Preserving Mappings}
\label{sec:pde}

We want to find a mapping $u\in C^{0,1}(\Omega,\Omega')$ such that  
\beq
\label{eq:measure-preserving}
\det(\nabla u(x)) =1\qquad \text{for any } x\in\Omega\;.
\eeq
When $\Omega=B(0,1)$ is the Euclidean unit ball and $\Omega'$ is star-shaped, \citet[Theorem~5.4]{Fonseca-Parry-1992} construct a mapping $u\in W^{1,\infty}(B(0,1),\Omega')$. They first construct $a\in W^{1,\infty}(B(0,1), \Omega')$ such that for any $x\in B(0,1)$, $\det(D_x a)=\lambda(x)$ for some $\lambda(x)>0$. Then using results of \citet{Dacorogna-Moser-1990}, a map $b\in C^{1,\alpha}(B(0,1), B(0,1))$ with $0<\alpha<1$ is constructed such that 
\begin{align*}
\det(D_x b) &=\lambda(x) \qquad \text{in } B(0,1)\,,\\
b(x) & = x \qquad \text{on } \partial B(0,1) \;.
\end{align*}
Define a mapping from $\Omega'$ to $B(0,1)$ by setting $z(x) = b\, \circ\, a^{-1} (x)$ for $x\in \Omega'$. Because 
\[
D_x z = (D_{a^{-1}(x)} b)  (D_{a^{-1}(x)} a )^{-1}\,, 
\]
we get that 
\[
\det(D_x z) = \lambda(a^{-1}(x)) \frac{1}{\lambda(a^{-1}(x))} = 1 \;.
\]
We get the desired embedding by setting $u = z^{-1}$. 

We show a more general approach to construct such mappings by using divergence-free (incompressible) vector fields. 
Consider velocity field 
\[
v(t,x):[0,1]\times\Real^n\ra\Real^n \,,
\]
and the ordinary differential equation
\[
\begin{cases}
\partial_t z(t) = v(t,z(t))\,, &\quad t > 0\\
z(0) = z_0 \,,
\end{cases}
\]
where $\partial_t$ denotes partial derivative with respect to time. Assume that $v$ is Lipschitz with respect to the spatial variable with Lipschitz constant $L$, uniformly with respect to the time variable. Under these conditions, by Picard--Lindel\"{o}f theorem~\citep{Hartman-2002} the above ODE has a unique and Lipschitz solution. Further, the classical flow of $v$, i.e. the flow $\Phi(t,x):[0,1]\times\Real^n\ra\Real^n$ satisfying 
\[
\begin{cases}
\partial_t \Phi(t,x) = v(t,\Phi(t,x))\,, &\quad t > 0\\
\Phi(0,x) = x \,,
\end{cases}
\]
is biLipschitz: for all $x_1,x_2\in\Real^n$,
\beq
\label{eq:PhiLip}
e^{-L t} \abs{x_1 - x_2} \le \abs{\Phi(t,x_1) - \Phi(t,x_2)} \le e^{L t} \abs{x_1 - x_2} \;.
\eeq

We want to choose $v$ and $\Phi$ such that $\Omega_1 = \Omega'$ and 
\[
u(x) = \Phi(1,x)
\]
is a solution for \eqref{eq:measure-preserving}. Let $\Omega_t$ be a subset of $\Real^n$ such that $\Omega_0=\Omega$ and $\Omega_t = \Phi(t,\Omega)$. To ensure that the flow is volume-preserving, we require that $v$ is divergence-free
\beq
\label{eq:divergence-free}
\begin{cases}
\DIV v(t,.) = 0 &\quad \text{in } \Omega_t\\
v(t,x) = f_t(x) &\quad \text{on } \partial\Omega_t \\
\end{cases}
\eeq
where boundary values $(f_t)$ are such that $\Omega_1=\Omega'$. By Divergence Theorem, we must have
\beq
\label{eq:fn}
\int_{\partial \Omega_t} f_t^\top \widehat n dS = 0\,,
\eeq
where $\widehat n$ is the outward pointing unit normal field of the boundary $\partial \Omega_t$. By chain rule, 
\[
\partial_t \nabla \Phi(t,x) = \nabla v(t,\Phi(t,x)) \nabla \Phi(t,x) \;.
\]
We know that if $\Psi(t)' = A(t) \Phi(t)$, then $(\det \Psi)' = \trace(A) \det \Psi$. Thus, 
\[
\partial_t \det \nabla \Phi(t,x) = (\det \nabla \Phi(t,x)) \DIV(v(t,\Phi(t,x))) = 0 \;.
\]
Thus, $\det \nabla \Phi(t,x) = \det \nabla \Phi(0,x) = 1$ for any $t\in [0,1]$. Thus, $u(x) = \Phi(1,x)$ is a solution for \eqref{eq:measure-preserving}. By \eqref{eq:PhiLip}, $\Phi$ (and hence $u$) inherits the smoothness of $v$. 
Thus it only remains to show a Lipschitz solution for \eqref{eq:divergence-free}. 

Problem~\eqref{eq:divergence-free} does not necessarily have a unique solution. Here we describe a solution based on a reduction to a Dirichlet problem. Assume $v(t,.) = \nabla h_t$ for some potential $h_t:\Real^n\ra\Real$. Let $c_t(\Real^n; \Real)$ be such that $\nabla c_t = f_t$ on $\partial \Omega_t$. Thus we want to solve the following Dirichlet problem for Laplace's equation
\beq
\label{eq:laplace}
\begin{cases}
\nabla^2 h_t = 0 &\quad \text{in } \Omega_t\\
h_t = c_t &\quad \text{on } \partial\Omega_t \\
\end{cases}
\eeq
Extend $c_t$ to the whole $\Omega_t$ and let $m_t = \nabla^2 c_t$. Solve
\beq
\label{eq:laplace2}
\begin{cases}
\nabla^2 w_t = m_t &\quad \text{in } \Omega_t\\
w_t = 0 &\quad \text{on } \partial\Omega_t \\
\end{cases}
\eeq
Then $h_t = c_t - w_t$ is a solution for \eqref{eq:laplace}. This holds because $h_t=0$ on $\partial\Omega_t$ and $\nabla^2 h_t = \nabla^2 c_t - \nabla^2 w_t = 0$ in $\Omega_t$. Thus 
\[
v(t,.) = \nabla h_t = \nabla c_t - \nabla w_t
\]
is a solution for \eqref{eq:divergence-free}. Lipschitzness of $v(t,.)$ can be shown by bounding second derivatives of $w_t$ and $c_t$. Regularity results for the solution of the Dirichlet problem for Laplace's equation exist. For example, we can use results of \citet[Chapter~6]{Gilbarg-Trudinger-2001} to show H\"{o}lder continuity of the second derivatives of solution of \eqref{eq:laplace2} given that $\partial\Omega_t$ and $m_t$ are sufficiently smooth.